\documentclass[aps,prl,twocolumn,superscriptaddress,showpacs]{revtex4}

\usepackage{graphicx}
\usepackage{epsfig}
\usepackage{ulem}
\usepackage{color}

\begin{document}


\title{Resistive switching induced by electronic avalanche breakdown in GaTa$_4$Se$_{8-x}$Te$_x$ narrow gap Mott Insulators}



\author{V. Guiot}
\affiliation{Institut des Mat$\acute{e}$riaux Jean Rouxel (IMN), Universit$\acute{e}$ de Nantes, CNRS, 2 rue de la Houssini$\grave{e}$re, BP32229, 44322 Nantes, France}

\author{L. Cario}
\email[]{Laurent.Cario@cnrs-imn.fr}
\affiliation{Institut des Mat$\acute{e}$riaux Jean Rouxel (IMN), Universit$\acute{e}$ de Nantes, CNRS, 2 rue de la Houssini$\grave{e}$re, BP32229, 44322 Nantes, France}

\author{E. Janod}
\affiliation{Institut des Mat$\acute{e}$riaux Jean Rouxel (IMN), Universit$\acute{e}$ de Nantes, CNRS, 2 rue de la Houssini$\grave{e}$re, BP32229, 44322 Nantes, France}

\author{B. Corraze}
\affiliation{Institut des Mat$\acute{e}$riaux Jean Rouxel (IMN), Universit$\acute{e}$ de Nantes, CNRS, 2 rue de la Houssini$\grave{e}$re, BP32229, 44322 Nantes, France}

\author{V. Ta Phuoc}
\affiliation{GREMAN, CNRS UMR 7347\\Universit\'{e} F.Rabelais. UFR Sciences - Parc de
Grandmont. 37200 Tours- France} 

\author{M. Rozenberg}
\affiliation{Laboratoire de Physique des Solides, CNRS UMR 8502, Universit$\acute{e}$ Paris Sud, B$\hat{a}$t 510, 91405 Orsay, France}

\author{P. Stoliar}
\affiliation{Laboratoire de Physique des Solides, CNRS UMR 8502, Universit$\acute{e}$ Paris Sud, B$\hat{a}$t 510, 91405 Orsay, France}

\author{T. Cren}
\affiliation{Institut des Nanosciences de Paris, Universit$\acute{e}$ Pierre et Marie Curie, CNRS UMR 7588, 4 place Jussieu, F-75005 Paris, France}

\author{D. Roditchev}
\affiliation{Institut des Nanosciences de Paris, Universit$\acute{e}$ Pierre et Marie Curie, CNRS UMR 7588, 4 place Jussieu, F-75005 Paris, France}


\date{\today}


\maketitle

\textbf{
Mott transitions induced by strong electric fields are receiving a growing interest. Recent theoretical proposals have focused on the Zener dielectric breakdown in Mott insulators, however experimental studies are still too scarce to conclude about the mechanism. Here we report a study of the dielectric breakdown in the narrow gap Mott insulators GaTa$_4$Se$_{8-x}$Te$_x$. We find that the I-V characteristics and the magnitude of the threshold electric field (E$_{th}$) do not correspond to a Zener breakdown, but rather to an avalanche breakdown. E$_{th}$ increases as a power law of the Mott Hubbard gap (E$_g$), in surprising agreement with the universal law E$_{th}$ $\propto$E$_g$$^{2.5}$ reported for avalanche breakdown in semiconductors. However, the delay time for the avalanche that we observe in Mott insulators is over three orders of magnitude longer than in conventional semiconductors. Our results suggest that the electric field induces local insulator-to-metal Mott transitions that create conductive domains which grow to form filamentary paths across the sample.}

In many materials with partially filled $d$ bands, the electron motion is frozen as a result of strong repulsive interactions, which drive the system into a Mott insulating state \cite{Mott, Imada}. Introduction of carriers in a Mott insulator often leads to exotic properties, such as high-Tc superconductivity in cuprates \cite{Berdnoz86} and colossal magnetoresistance in manganites \cite{Jin94}. A Mott insulator can also be collapsed into a correlated metal by applying pressure. Notable examples are V$_{2}$O$_3$ \cite{Limelette03} and  organics \cite{Kawaga05}. Another important type of Mott transition, which is receiving a great deal of attention, is the dielectric breakdown caused by strong electric fields. Many recent theoretical studies have focused on the Zener mechanism of breakdown for Mott insulators. For instance, calculations were performed in 1D Hubbard chains using exact diagonalization \cite{Oka05, Oka10, Oka03, Oka05b} and time-dependent density matrix renormalization group \cite{Heidrich10}, or in the limit of large dimensions using dynamical mean field theory \cite{Eckstein10}. These theoretical studies predicted a similar non-linear behavior in the current-voltage characteristics, and the existence of a threshold field (E$_{th}$) beyond which a field induced metal appears.  In the latter case, where the Mott transition occurs beyond a critical value of the Coulomb interaction which is of similar magnitude as the bandwidth, it was found that E$_{th}$ is proportional to $\Delta^{2}/v_F$, with $\Delta$ the charge gap, and $v_F$ the Fermi velocity. This implies a large threshold field of the order of 10$^{6}$-10$^{7}$ V/cm, which is substantially larger than the experimental values observed in dielectric breakdown \cite{Taguchi00}. Surprisingly, compared to this significant theoretical effort, experimental studies of Mott insulators under strong electric fields remain quite scarce. Taguchi et al. observed a dielectric breakdown in the quasi-one-dimensional Mott insulators Sr$_{2}$CuO$_{3}$ and SrCuO$_{2}$ \cite{Taguchi00}. More recently a comparable dielectric breakdown was observed in the family of chalcogenide Mott insulators AM$_{4}$Q$_{8}$ (A = Ga, Ge; M = V, Nb, Ta, Mo; Q = S, Se) \cite{Vaju08, Vaju08bis, Dubost09, Cario10, Souchier11}. However in both cases E$_{th}$ is of the order of 10$^{2}$-10$^{4}$ V/cm which does not match with theoretical predictions for Zener breakdown. This strong discrepancy raises therefore the question of the actual physical origin of the observed breakdown. Here we focus our study on the chalcogenide Mott insulators AM$_4$Q$_8$ and show that the dielectric breakdown is consistent with an electric field induced avalanche phenomenon rather than Zener breakdown. Our work reveals that the I(V) curve, the magnitude of E$_{th}$  and the dependence of  E$_{th}$  versus the Mott-Hubbard gap are those expected for an avalanche breakdown,  qualitatively similar to what is observed in semiconductors. However, our results also reveal significant differences, for instance, a delay time for the avalanche that is at least three orders of magnitude longer in the Mott insulators AM${_4}$Q${_8}$ compared to  the case of conventional band insulators.

\begin{center}
\textbf{RESULTS}
\end{center}

\begin{figure}[t]
\centerline{\includegraphics[width=85 mm]{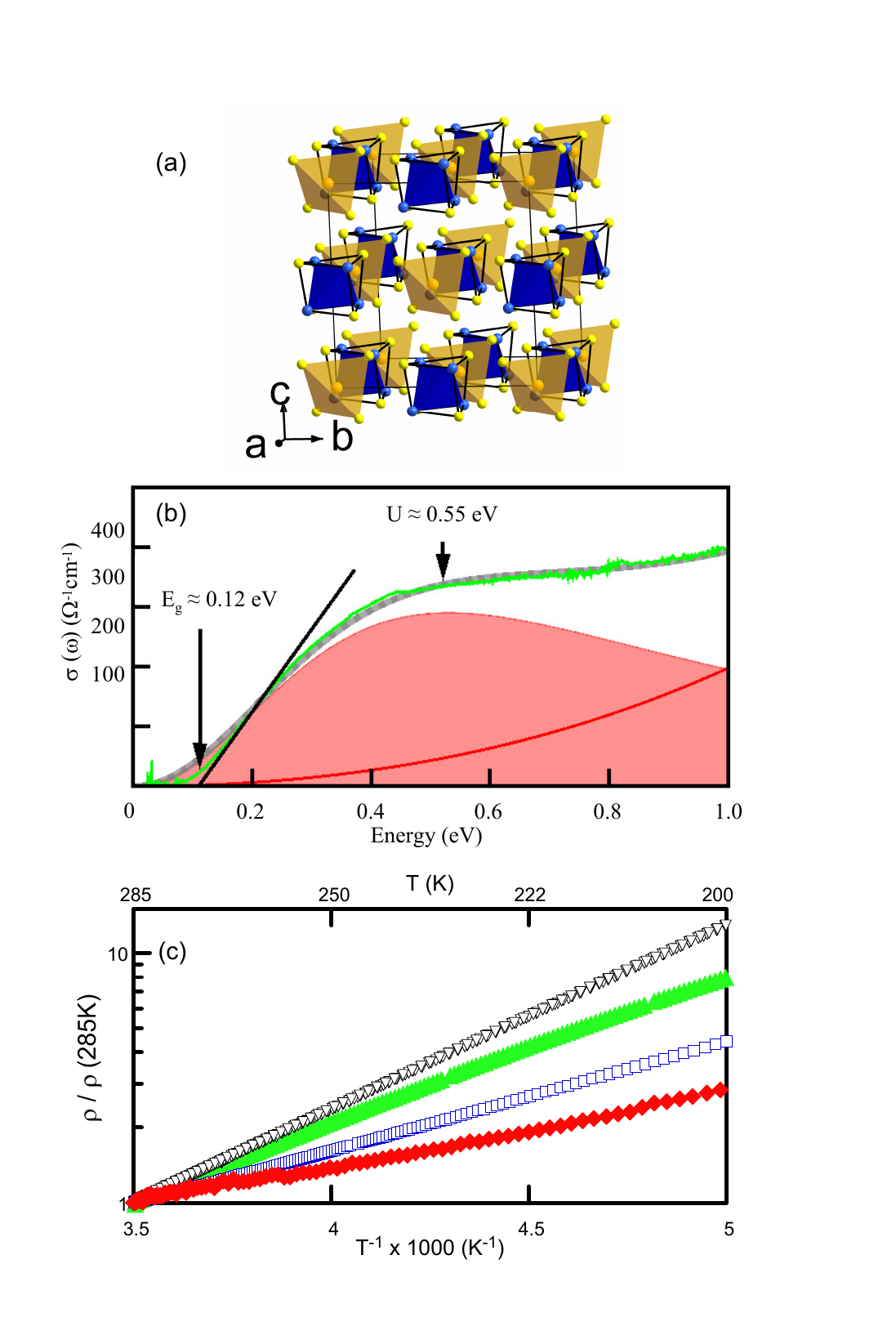}}
\caption{\textbf{Resistivity and optical conductivity measurements.} (a) representation of the lacunar spinel structure of the AM$_4$Q$_8$ compounds. Blue, orange and Yellow filled circle represents M, A and Q atoms respectively. The M$_4$ tetrahedral cluster and the AQ$_4$ tetrahedra are highlighted in blue and orange, respectively (b) experimental optical conductivity of GaTa$_4$Se$_8$ (green line) showing that the optical conductivity is that expected for a Mott Insulator with an Hubbard energy U= 0.55eV and a gap close to 0.12 eV. The gray solid line corresponds to the fit of optical conductivity spectrum using two contributions : a broad mid-infrared band (red shaded area) and high-frequencies contributions (thick solid red line). (c) Resistivity as a function of inverse temperature for GaTa$_4$Se$_{8}$ (filled green triangle), GaTa$_4$Se$_{4}$Te$_{4}$ (black open triangle), GaTa$_4$Se$_{2}$Te$_{6}$ (blue open square), and GaTa$_4$Se$_{1.5}$Te$_{6.5}$ (filled red square).} 
\label{fig:figure1}
\end{figure}

\begin{center}
\textbf{Mott-Hubbard gap evolution in the GaTa${_4}$Se$_{8-x}$Te$_{x}$ compounds}
\end{center}
The lacunar spinel compounds AM${_4}$Q${_8}$ , with tetrahedral transition metal clusters (see Fig. \ref{fig:figure1}) \cite{Yaich84}, represents an interesting family of narrow gap Mott insulators. The low Mott gap value makes these compounds highly sensitive to doping or pressure. Doping of GaV$_{4}$S$_{8}$ by Ti leads, for example, to a filling control Insulator to Metal  Transition (IMT) with the emergence of a ferromagnetic half metal \cite{Vaju08ter, Dorolti10}. On the other hand, an external pressure induces a bandwidth controlled IMT and superconductivity in GaTa$_{4}$Se$_{8}$ \cite{Abd04, Pocha05}. Recently, we have reported that electric field leads to a striking volatile IMT or dielectric breakdown in GaTa$_4$Se$_8$ \cite{Vaju08, Vaju08bis, Dubost09, Cario10} which is qualitatively different from the non-volatile resistive switching based on electrochemical effects reported so far in many transition metal oxides \cite{Waser07, Rozenberg11, Cario10}. 

In order to study the origin of this dielectric breakdown and to explore its relationship with the Mott Hubbard gap we have investigated the electric field effects in the series of compounds GaTa${_4}$Se$_{8-x}$Te$_{x}$ (x=0, 4.0, 6.0, 6.5). These compounds were synthesized as described elsewhere \cite{Guiot11}. Optical conductivity and resistivity measurements were subsequently undertaken to evaluate the gap evolution in this series. The room temperature optical conductivity of GaTa$_{4}$Se$_{8}$, shown in Fig. \ref{fig:figure1}b, displays all the expected features of a narrow gap Mott Insulator with a Hubbard energy (U) of the order of 0.55 eV {\cite{TaPhuoc12}}. The low bias resistivities $\rho$(T) measured with a standard four probes technique for the series of compounds GaTa$_{4}$Se$_{8-x}$Te$_{x}$ exhibit a semiconducting-like behavior with a temperature dependence consistent with an activation law at high temperature (see Fig. \ref{fig:figure1}-c). The systematic change of the activation energy observed in Fig. \ref{fig:figure1} suggests that the substitution of Se per Te permits the bandwidth-control of the Mott Hubbard gap (from 100 to 300 meV) within the GaTa$_{4}$Se$_{8-x}$Te$_{x}$ series of compounds, thus enabling the study of the threshold field dependence with the size of the gap.

\begin{center}
\textbf{Dielectric breakdown experiments in the GaTa${_4}$Se$_{8-x}$Te$_{x}$ Mott insulators}
\end{center}

Dielectric breakdown was systematically studied in this series of narrow gap Mott insulators using a procedure described in the Methods section. Figure 2 shows that these compounds undergo a sudden volatile dielectric breakdown above a certain threshold electric field ($\approx$1-10 kV/cm). As an example, Fig. 2c depicts the typical temporal evolution of the voltage V$_S$(t) across a GaTa$_{4}$Se$_{8}$ crystal while a series of short (10 to 70 $\mu$s) voltage pulses (V$_{pulse}$) are applied.  When the electric field applied to the crystal exceeds a threshold field of $\approx$ 2.5 kV/cm the crystal systematically undergoes a fast transition to a low resistance state (i.e. from 70 k$\Omega$  to 5 k$\Omega$ ) called a dielectric breakdown or resistive switching. This transition is volatile and perfectly reproducible. In fact, after the pulse application the low-bias resistance returns to its original value. The delay t$_{delay}$ between the application of the pulse  and the dielectric breakdown shifts to smaller values with increasing applied voltage. The voltage drop on the sample V$_{S}$ in the breakdown state is always $V_S$ $\approx$ 9 V (or $E_{th}$ $\approx$ 2.5 kV/cm) whatever the voltage applied to the sample before the resistive switching. On the other hand, the current flowing through the sample after the dielectric breakdown shows a highly non-ohmic  behavior in the transited state in both, pulse or dc measurements (see Fig. 2-b). Thus, we infer that E$_{th}$ (i.e. V$_{th}$/$d$ where $d$ is the distance between electrodes) corresponds to the breakdown field for the limiting case t$_{delay}$ $\rightarrow$ $\infty$.  The same qualitative behavior was observed for GaTa${_4}$Se$_{1.5}$Te$_{6.5}$ (see Fig. 2a), all other GaTa${_4}$Se$_{8-x}$Te$_{x}$ compounds, and more generally for all AM$_4$Q$_8$ (A = Ga, Ge; M = V, Nb, Ta, Mo; Q = S, Se) Mott insulator compounds \cite{Cario10}.

\begin{figure}[t]
\centerline{\includegraphics[width=85 mm]{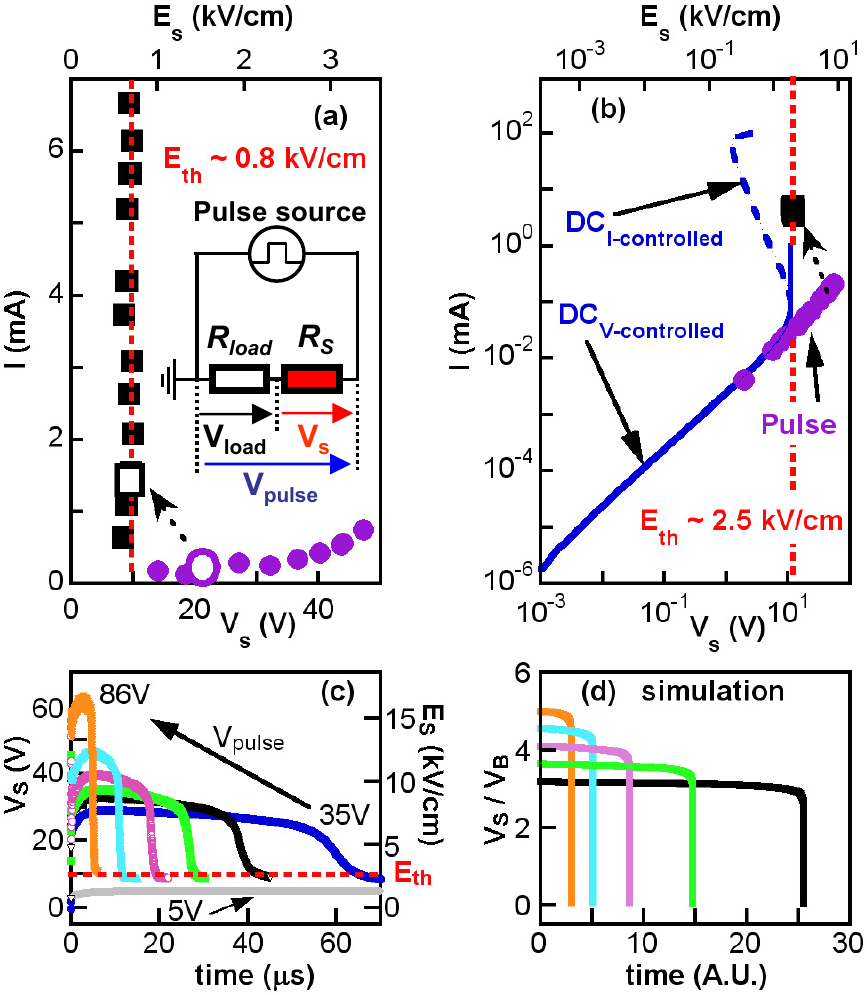}}
\caption{\textbf{Resistive switching experiments and modeling.} (a)I-V curves extracted from pulse (open circles and squares) experiments for a crystal of GaTa$_4$Se$_{1.5}$Te$_{6.5}$. The circuit used to apply voltage pulses is shown in inset. (b) I-V curves extracted from pulse (open circles and squares) or DC experiments (plain or dashed lines) for a crystal of GaTa$_4$Se$_8$. (c) Evolution of the voltage V$_s$(t) across a crystal of GaTa$_4$Se$_8$ when short voltage pulses V$_pulse$ (10-70 $\mu$s) of increasing magnitude are applied to the circuit. For example, V$_pulse$ = 5V , 35V, 40V, 44V, 50V, 60V and 86V for the gray, blue, black, green, pink, light blue and orange curves, respectively. These curves show a clear evolution of the time (t$_{delay}$) at which occurs the transition to a low resistance state (resistive switching) as a function of V$_{pulse}$. In (a) and (b) the I-V curves are extracted from V$_s$(t) curves similar to the one shown in (c) and from the corresponding I(t) curves not shown here. In (a) and (b) open circles correspond to the I-V values measured before the resistive switching and open squares correspond to the I-V values measured after the resistive switching. (d) Results of mathematical modeling of V$_s$(t) using a one dimensional resistive network. The time at which occurs the resistive switching decreases as the applied voltage to the resistor network is increased from the black to the orange curve.}
\label{fig:figure2}
\end{figure}

\begin{center}
\textbf{Dielectric breakdown mechanism in the GaTa${_4}$Se$_{8-x}$Te$_{x}$ Mott insulators}
\end{center}

For conventional semiconductors the dielectric breakdown is well documented and different mechanisms are proposed such as Zener tunneling \cite{Zener34} or avalanche process \cite{Levinshtein05}. The threshold field observed in the AM$_4$Q$_8$ compounds is at least 3 orders of magnitude smaller than the threshold field of the order of 10$^{6}$ - 10$^{7}$V/cm expected for Zener breakdown in Mott insulators \cite{Oka05}. This definitively rules out a Zener breakdown mechanism. Conversely, the magnitude of the threshold field in AM$_4$Q$_8$ Mott insulators compares well with the threshold field values of 1 and 40 kV/cm observed for avalanche breakdowns in narrow gap semiconductors InSb and InAs, respectively \cite{Hudgins03, Hudgins03b}. The I(V) characteristics observed in the AM$_4$Q$_8$ compounds and in semiconductors like InSb, InAs, Si or Ge are also very similar \cite{Levinshtein05}. Indeed, all these compounds exhibit a steep increase of the current at the threshold field, which is almost vertical. In narrow gap semiconductors this is related to the impact- ionization process. An avalanche breakdown indeed occurs when the electrons accelerated by the applied voltage can gain enough energy to ionize by impact other electrons above the band gap and generate electron-hole pairs \cite{Levinshtein05}. When the creation of electron-hole pair by impact-ionization overcompensates the electron-hole recombination process (Auger Thermal process) a multiplication of carriers occurs and induces a dielectric breakdown with a steep increase of the current at the threshold field. For high current density the avalanche breakdown is characterised by a negative differential resistance (NDR) in the current-voltage characteristic which is the consequence of the contribution of holes to the field distribution along the sample \cite{Levinshtein05}. As shown in Fig. 2, such a NDR effect is observed on the I-V curve of GaTa$_4$Se$_8$ measured in DC mode using a current source.  Again, this strongly supports an avalanche mechanism as the origin of the dielectric breakdown. 

\begin{center}
\textbf{Dependence of the avalanche threshold field in Mott insulators and classical semiconductors}
\end{center}

For semiconductors the avalanche threshold field varies as a function of a power law of the band gap \cite{Levinshtein05} and follows the universal law E$_{th}$ $\propto$ Eg$^{2.5}$ \cite{Hudgins03, Hudgins03b}. For comparison Fig. 3 displays the variation of the threshold field as a function of the Mott-Hubbard gap for all the AM$_4$Q$_8$ compounds. This figure reveals a remarkable power law dependence of the threshold fields with the gap E$_{th}$ $\propto$ Eg$^{2.5}$ (see blue curve in fig 3). The inset of Fig. 3 compares also the threshold fields versus gap dependence for the AM$_4$Q$_8$ compounds and classical semiconductors. This comparison reveals that for all Mott and band insulators or semiconductors the threshold fields line up along a universal law E$_{th}$ $\propto$ Eg$^{2.5}$. This is again strong evidence that the breakdown observed in the Mott Insulators AM$_4$Q$_8$ originates from an avalanche phenomenon.

\begin{figure}[t]
\centerline{\includegraphics[width=85 mm]{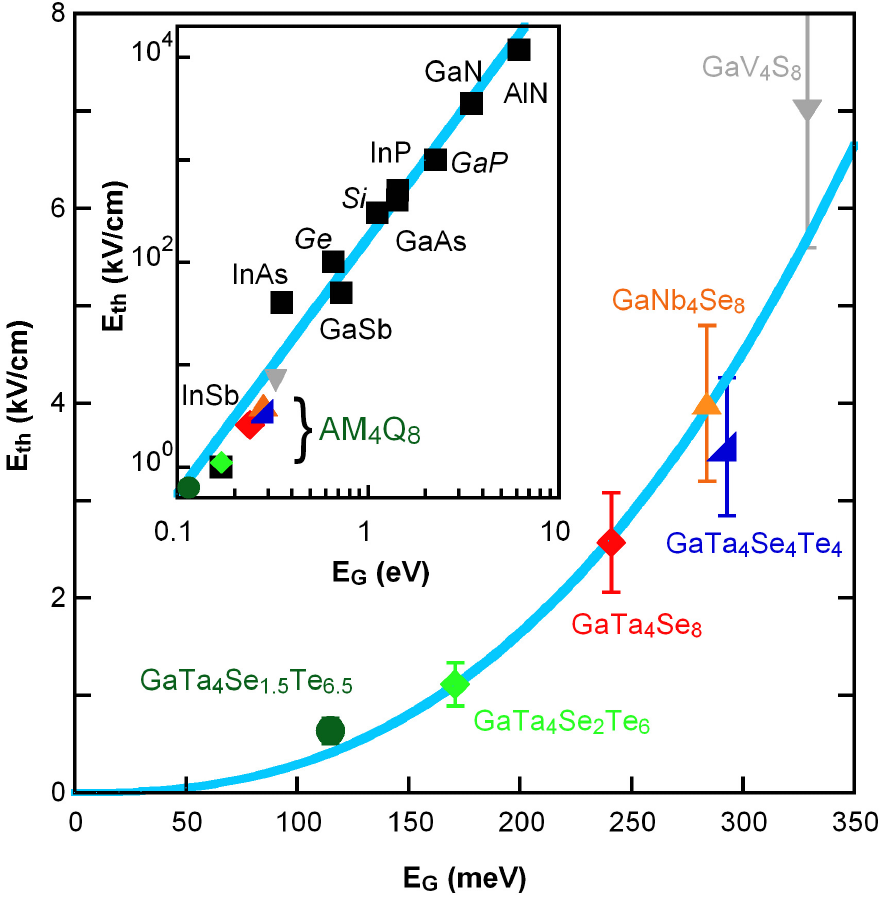}}
\caption{\textbf{Dependence of E$_{th}$ in Mott insulators and semiconductors.} Threshold Electric field (inducing avalanche breakdown) as a function of the Mott gap for various AM$_4$Q$_8$ compounds. The solid blue curve correspond to a power law dependence E$_{th}$ $\propto$ Eg$^{2.5}$. Inset : comparison of the threshold fields versus gap dependence for the AM$_4$Q$_8$ compounds and for classical semiconductors. The solid blue line displays the universal law E$_{th}$[kV/cm] = 173 (E$_g$[eV])$^{2.5}$ observed for semiconductors.}
\label{fig:figure3}
\end{figure}

\begin{center}
\textbf{DISCUSSION}
\end{center}

This scaling of the Mott- and band-insulators along the same universal law E$_{th}$ = 1.73 E$_g$$^{2.5}$ indicates that the threshold field in the avalanche breakdown depends only on the gap whatever its origin and nature. This is rather surprising as the excitation gap in Mott insulators arises from electron-electron repulsions, (i.e. from strong correlation effects), in stark contrast with the case of conventional semiconductors.
Despite this similarity, as we mentioned before, the main qualitative difference between the electric breakdown in semiconductors and Mott insulators is the significantly longer t$_{delay}$ of the latter. The origin of the difference may be due to the different coupling to the lattice degrees of freedom. The Mott transition is indeed accompanied of a compressibility anomaly \cite{Kotliar02, Hassan05} which leads to a large volume change \cite{mcwhanV2O3}.
In fact, the giant electromechanical coupling observed during an STM studies by applying electric pulses between the tip and a cleaved GaTa$_4$Se$_8$ crystal surface supports this scenario \cite{Dubost09}. Therefore, the strong coupling of the electronic and lattice degrees of freedom is likely to slow down significantly the velocity of the avalanche propagation. Interestingly, Fig. 2 shows that the dielectric breakdown in the AM$_4$Q$_8$ Mott insulators occurs with a t$_{delay}$ in the $\mu$s range (from a few $\mu$s to the ms) for typical sample lengths $d$=20-40 $\mu$m. In contrast, in semiconductors the $t_{delay}$ lies in the ps to the ns range, that is, several orders of magnitude faster \cite{Levinshtein05}. Figure 4 displays the avalanche velocity $v=d/t_{delay}$ as a function of excess electric field E-E$_{th}$ where E is the electric field applied on the sample. This figure shows that $v$ varies as (E-E$_{th}$)$^3$. This remarkable power law dependence as well as the magnitude of the velocity in $m/s$ range recall a phase growth mechanism \cite{Willnecker90} or the propagation of dislocations \cite{Johnston59, Messerschmidt10}. This analysis therefore suggests that the excess electric field triggers the local growth of conducting  domains where a Mott transition has taken place, which eventually form filamentary paths across the sample as the avalanche breakdown occurs.

\begin{figure}[t]
\centerline{\includegraphics[width=55 mm]{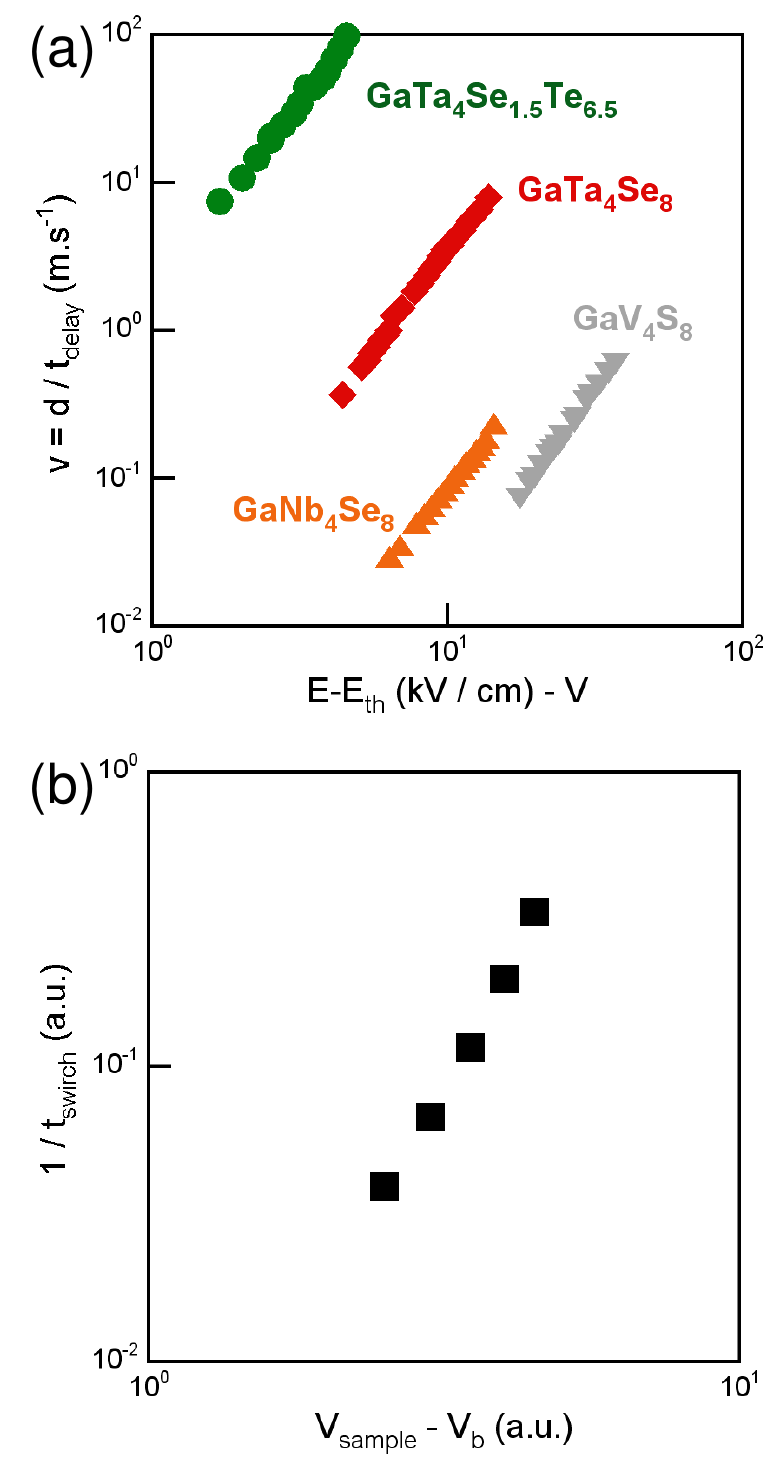}}
\caption{\textbf{Avalanche velocity in Mott Insulators.} (a) Avalanche velocity $v=d/t_{delay}$ for various AM$_4$Q$_8$ compounds as a function of excess electric field E-E$_{th}$ where E is the electric field applied on the sample. (b) velocity of the transition 1/t$_{delay}$ extracted from the mathematical modeling of the resistive switching shown in figure 2d.}
\end{figure}

A mathematical modeling of the previous scenario was undertaken to describe the experimental data.  
This model considers a one dimensional resistive network where each cell
may be in one of two states, insulator (I) or metal (M), characterised
by their corresponding resistance $r_I$ and $r_M$, and their energy $E_I$ and $E_M$. This model also assumes that the two states are separated by an energy barrier of height $E_B$-$E_I$. The energy landscape used for the calculation is shown in Supplementary Figure S1. Initially, all cells are in the insulating state, since the system is a Mott insulator. An external voltage pulse $V_{pulse}$ is then applied on the resistor network and the local voltage drops on the cells are computed. The cells may undergo a transition from state I to M by a thermal activation process.The effective activation energy barrier, for an I to M transition of the $i$-th cell under the action of the applied voltage, is(E$_B$ - E$_I$ - $\Delta$V$_i$), where $\Delta$V$_i$ is the computed voltage drop at the cell. These assumptions lead to non-linear differential equation for the rate of cell transitions per unit time. Further details of the mathematical expressions are provided in the Methods section.
The solution of the differential equation is easily obtained by standard numerical integration.
The results are shown in Fig.2(d). We observe that this simple model qualitatively captures the behavior
of $t_{delay}$ of the avalanche mechanism. Initially, the conductive path modeled by the resistive network is highly resistive, so there is a small voltage drop on each cell, and the rate of transitions is low. Successive transitions,
however, lower the resistance of the path, as r$_M$ $<<$ r$_I$, therefore increasing the current and also the
voltage drop $\Delta$V$_i$ in the cells that still remain insulating. The increased $\Delta$V$_i$ lowers the local
energy activation barriers, thus augmenting the transition rates. A larger fraction of cells undergo a transition
and, eventually, a runaway process take place were the rest of the cells rapidly completes their transitions.
Interestingly, this simple model also qualitatively accounts for the field dependence of the velocity of the transition as shown in Supplementary Figure S2 and in Fig. 4. However the power law behavior predicted by this simple model is parameter dependent, i.e. is non universal. 

To conclude we have reported for the first time an avalanche breakdown in the narrow gap Mott insulators AM$_4$Q$_8$. The phenomenology of the breakdown (I(V) characteristic) and the magnitude of the threshold electric field are comparable with those reported for avalanche breakdowns in semiconductors. Moreover the threshold field dependence with the width of the Mott Hubbard gap shows a surprising agreement with the universal law E$_{th}$ = 1.73 Eg$^{2.5}$ reported for semiconductors, despite the fact that the AM$_4$Q$_8$ are strongly correlated systems. Our experimental and theoretical investigations suggest that, unlike semiconductors, avalanche breakdown in Mott insulators is associated to the growth of conducting filamentary paths of a metallic metastable phase within the sample. The strong coupling of electronic and lattice degrees of freedom due to compressibility anomalies may account for the reduced avalanche speed observed in the Mott insulators compared to conventional band insulators. Finally, our work suggests that avalanche breakdown may be a general feature of narrow gap Mott insulators. 

\begin{center}
\textbf{METHODS}
\end{center}
\begin{center}
\textbf{Sample preparation}
\end{center}
Single crystals of GaTa$_4$Se$_{8-y}$Te$_y$ with y ranging from 0 to 6.5 were prepared using stoichiometric mixtures of elemental gallium, tantalum, selenium and tellurium (purities $\geq$ 99,5 $\%$) \cite{Guiot11}. These mixtures were loaded in evacuated sealed quartz ampules of about 10 cm and heated up at 300$^{\circ}$C/h to 1000$^{\circ}$C and hold at this temperature during 160h. The quartz tubes were subsequently cooled down, first slowly to 800$^{\circ}$C at 1$^{\circ}$C/h and then faster (300$^{\circ}$C$/$h) to room temperature. These syntheses yielded black powders containing a high yield of GaTa$_4$Se$_{8-y}$Te$_y$ metallic-gray cubic and tetrahedral crystals (typical size $\leq$ 300 $\mu$m) along with a small amount of impurity phases (e.g. GaTe, TaSe$_2$). Large amounts of tetrahedral or cubic-shaped crystals  of all other members of the AM$_4$Q$_8$ family have been obtained using a similar procedure. 

\begin{center}
\textbf{Transport measurements}
\end{center}
The AM$_4$Q$_8$ crystals used for transport measurements were contacted using 17 $\mu$m gold wires and carbon paste (Electrodag PR-406), and then annealed in vacuum at 150$^{\circ}$C during 30 minutes. The low-bias resistance of the AM$_4$Q$_8$ was measured using a source-measure unit (either a Keithley 236 or a Keithley 6430) by a standard two- or four-probe technique. We checked that the contact resistances were much smaller than the sample resistance. Voltage pulses were applied using an Agilent 8114A. During the pulse, the voltage and current across the sample were measured with a Tektronix DPO3034 oscilloscope associated with a IeS-ISSD210 differential probe, using the simple circuit described in the inset of Fig 2-a. 

\begin{center} 
\textbf{Mathematical Modeling}
\end{center}

In order to obtain Figures 2d and 4b we numerically integrated
\begin{equation}
\frac{dx}{dt}=\nu\cdot\exp\left[-\frac{E_{B}-E_{I}-\triangle V_{i}}{k_{B}T}\right]
\end{equation}
by using 4th order Runge-Kutta method (time step = 0.001). Here $x$
is the fraction of cells in metallic state. E$_{B}$, E$_{I}$, and $\triangle$V$_{i}$ are defined as displayed in Supplementary Figure S1. We set $\nu=0.001$, $k_{B}T=1$,
$E_{B}-E_{I}=1$ and
\begin{equation}
\triangle V_{i}=V_{pulse}\cdot\frac{r_{I}}{r_{I}\cdot\left(1-x\right)+r_{M}\cdot x+R_{L}}.
\end{equation}
For Figure 2d we simulated $V_{pulse}=$ 2.5, 3, 3.5, 4, 4.5, 5 and
5.5. Figure 4b presents the switching times for the 5 higher values.
The 1D resistivities were set as $10\cdot r_{M}=R_{L}$ and $r_{I}=10\cdot R_{L}$.
$V_{S}$ was obtained as:
\begin{equation}
V_{S}=V_{pulse}\cdot\frac{r_{I}\cdot\left(1-x\right)+r_{M}\cdot x}{r_{I}\cdot\left(1-x\right)+r_{M}\cdot x+R_{L}}.
\end{equation}
$t_{delay}^{-1}$ was calculated as the inverse of the number of Runge-Kutta
cycles needed for $x$ to arrive to 1.
Another output of this numerical simulation is presented in Supplementary Figure
S2, the evolution of $\frac{dx}{dt}$ in time.

\begin{center} 
\textbf{ACKNOWLEDGMENTS}
\end{center}

This work was supported by the French Agence Nationale de la Recherche through the funding of the NV-CER (ANR-05-JCJC-0123-01) and NanoMott (ANR-09-Blan-0154-01) projects. 

\begin{center} 
\textbf{Author Contributions}
\end{center}

L.C., E.J. and B.C. initiated the project. V.G., L.C., E.J. and B.C. performed the synthesis and the caracterisations of the crystals, and the resistive switching experiments. V.T.P. did the optical conductivity measurements. P.S. and M.R. did the modeling. L.C., E.J., B.C., V.G., V.T.P., D.R., T.C., P.S. and M.R discussed and wrote the article in the framework of the NanoMott project.

\begin{center} 
\textbf{Competing  financial interest}
\end{center}

the author declare no competing financial interest


\begin{thebibliography}{0}
\expandafter\ifx\csname natexlab\endcsname\relax\def\natexlab#1{#1}\fi
\expandafter\ifx\csname bibnamefont\endcsname\relax
  \def\bibnamefont#1{#1}\fi
\expandafter\ifx\csname bibfnamefont\endcsname\relax
  \def\bibfnamefont#1{#1}\fi
\expandafter\ifx\csname citenamefont\endcsname\relax
  \def\citenamefont#1{#1}\fi
\expandafter\ifx\csname url\endcsname\relax
  \def\url#1{\texttt{#1}}\fi
\expandafter\ifx\csname urlprefix\endcsname\relax\def\urlprefix{URL }\fi
\providecommand{\bibinfo}[2]{#2}
\providecommand{\eprint}[2][]{\url{#2}}

\end{thebibliography}


\begin{center}
\textbf{REFERENCES}
\end{center}
\begin{thebibliography}{}
\bibitem{Mott}Mott, N.F. Metal-Insulator Transitions. Taylor \& Francis, London (1990)
\bibitem{Imada}Imada, M. \textit{et al.} Metal-insulator transitions, \textit{Rev. Mod. Phys.} \textbf{70}, 1039-1263 (1998).
\bibitem{Berdnoz86}Bednorz, J. G. \textit{et al.} Possible high Tc superconductivity in the Ba-La-Cu-O system, \textit{Z. Physik B} \textbf{64}, 189-193 (1986).
\bibitem{Jin94}Jin, S. \textit{et al.} Thousandfold Change in Resistivity in Magnetoresistive La-Ca-Mn-O Films, \textit{Science} \textbf{264}, 413-415 (1994).
\bibitem{Limelette03}Limelette, P.  \textit{et al.} Universality and Critical Behavior at the Mott Transition, \textit{Science} \textbf{302}, 89-92 (2003).
\bibitem{Kawaga05}Kagawa, F. \textit{et al.} Unconventional critical behaviour in a quasi-two-dimensional organic conductor, \textit{Nature} \textbf{436}, 534-537 (2005).
\bibitem{Oka05}Oka, T. \textit{et al.} Ground-State Decay Rate for the Zener Breakdown in Band and Mott Insulators, \textit{Phys. Rev. Lett.} \textbf{95}, 137601 (2005).
\bibitem{Oka10}Oka, T. \textit{et al.} Dielectric breakdown in a Mott insulator: Many-body Schwinger-Landau-Zener mechanism studied with a generalized Bethe ansatz, \textit{Phys. Rev. B} \textbf{81}, 033103 (2010).
\bibitem{Oka03}Oka, T. \textit{et al.} Breakdown of a Mott Insulator: A Nonadiabatic Tunneling Mechanism, \textit{Phys. Rev. Lett.} \textbf{91}, 066406 (2003).
\bibitem{Oka05b}Oka, T. \textit{et al.} Nonlinear transport in a one-dimensional Mott insulator in strong electric fields. \textit{Physica B: Cond. Matter}  \textbf{359-361}, 759-761 (2005).
\bibitem{Heidrich10}Heidrich-Meisner, F. \textit{et al.} Nonequilibrium electronic transport in a one-dimensional Mott insulator, \textit{Phys. Rev. B} \textbf{82}, 205110 (2010).
\bibitem{Eckstein10}Eckstein, M. \textit{et al.} Dielectric Breakdown of Mott Insulators in Dynamical Mean-Field Theory, \textit{Phys. Rev. Lett.} \textbf{105}, 146404 (2010).
\bibitem{Taguchi00}Taguchi, Y. \textit{et al.} Dielectric breakdown of one-dimensional Mott insulators Sr$_2$CuO$_3$ and SrCuO$_2$, \textit{Phys. Rev. B} \textbf{62}, 7015 (2000).
\bibitem{Vaju08}Vaju, C. \textit{et al.} Electric-pulse-driven electronic phase separation, insulator-metal transition, and possible superconductivity in a Mott insulator. \textit{Adv. Mater.} \textbf{20}, 2760-2765 (2008).
\bibitem{Vaju08bis}Vaju, C. \textit{et al.} Electric-Pulse-Induced Resistive Switching and Possible Superconductivity in the Mott Insulator GaTa$_4$Se$_8$. \textit{Microelectronics Eng.} \textbf{85}, 2430-2433 (2008).
\bibitem{Dubost09}Dubost, V. \textit{et al.} Electric-Field-Assisted Nanostructuring of a Mott Insulator. \textit{Adv. Funct. Mater.} \textbf{19}, 2800-2804 (2009).
\bibitem{Cario10}Cario, L. \textit{et al.} Electric-Field-Induced Resistive Switching in a Family of Mott Insulators: Towards a New Class of RRAM Memories. \textit{Adv. Mater.} \textbf{22}, 5193-5197 (2010).
\bibitem{Souchier11}Souchier, E. \textit{et al.} First evidence of resistive switching in polycrystalline GaV$_4$S$_8$ thin layers. \textit{Phys. Status Solidi RRL} \textbf{5}, 53-55 (2011).
%
\bibitem{Yaich84}Ben Yaich, H. \textit{et al.} Nouveaux chalcog\'{e}nures et chalcohalog\'{e}nures \`{a} clusters t\'{e}tra\'{e}driques Nb$_4$ ou Ta$_4$, \textit{J. Less-Common Met.} \textbf{102}, 9-22 (1984).
\bibitem{Vaju08ter}Vaju, C. \textit{et al.} Metal-metal bonding and correlated metallic behavior in the new deficient spinel Ga$_{0.87}$Ti$_4$S$_8$. \textit{Chem. Mater.} \textbf{20}, 2382-2387 (2008).
\bibitem{Dorolti10}Dorolti, E. \textit{et al.} Half-Metallic Ferromagnetism and Large Negative Magnetoresistance in the New Lacunar Spinel GaTi$_3$VS$_8$. \textit{J. Amer. Chem. Soc.} \textbf{132}, 5704-5710 (2010).
\bibitem{Abd04}Abd-Elmeguid, M.M.  \textit{et al.}, Transition from Mott Insulator to Superconductor in GaNb$_4$Se$_8$ and GaTa$_4$Se$_8$ under High Pressure, \textit{Phys. Rev. Lett.} \textbf{93}, 126403 (2004).
\bibitem{Pocha05}Pocha, R. \textit{et al.} Crystal Structures, Electronic Properties, and Pressure-Induced Superconductivity of the Tetrahedral Cluster Compounds GaNb$_4$S$_8$, GaNb$_4$Se$_8$, and GaTa$_4$Se$_8$, \textit{J. Am. Chem. Soc.} \textbf{127}, 8732-8740 (2005).
\bibitem{Waser07}Waser, R. \textit{et al.} Nanoionics-based resistive switching memories, Nature Mater. \textbf{6}, 833-840 (2007); 
\bibitem{Rozenberg11}Rozenberg M.J. Resistive switching. \textit{Scholarpedia} \textbf{6(4)} 11414 (2011).
\bibitem{Guiot11}Guiot, V. \textit{et al.} Control of the electronic properties and resistive switching in the new series of Mott Insulators GaTa$_4$Se$_{8-y}$Te$_y$ (0$\leq$y$\leq$6.5). \textit{Chem. Mater.} \textbf{23}, 2611-2618 (2011).
\bibitem{TaPhuoc12}Ta Phuoc, V. \textit{et al.} Optical Conductivity Measurements of GaTa4Se8 Under High Pressure: Evidence of a Bandwidth-Controlled Insulator-to-Metal Mott Transition. \textit{Phys. Rev. Lett.} \textbf{110}, 037401 (2013).
\bibitem{Zener34}Zener, C. , A Theory of the Electrical Breakdown of Solid Dielectrics, \textit{Proc. Royal Soc. A} \textbf{145}, 523-529 (1934).
%
\bibitem{Levinshtein05}Levinshtein, M. \textit{et al.} Breakdown Phenomena in Semiconductors and Semiconductor Devices, World Scientific Publishing Co. Pte. Ltd., Singapore, (2005).
\bibitem{Hudgins03}Hudgins, J. L. \textit{et al.} An Assessment of Wide Bandgap Semiconductors for Power Devices, \textit{IEEE Transactions on Power Electronics} \textbf{18}, 907-914 (2003).
\bibitem{Hudgins03b}Hudgins, J. Wide and Narrow Bandgap Semiconductors for Power Electronics: A New Valuation, \textit{J. Electronic Mater.} \textbf{32}, 471-477 (2003). 
\bibitem{Kotliar02}Kotliar, G. \textit{et al.} Compressibility Divergence and the Finite Temperature Mott Transition, \textit{Phys. Rev. Lett.} \textbf{89}, 046401 (2002).
\bibitem{Hassan05}Hassan, S. R. \textit{et al.} Sound Velocity Anomaly at the Mott Transition: Application to Organic Conductors and V$_2$O$_3$, \textit{Phys. Rev. Lett.} \textbf{94}, 036402 (2005).
\bibitem{mcwhanV2O3}McWhan, D.B. \textit{et al.} Metal-Insulator Transitions in Pure and Doped V$_2$O$_3$, \textit{Phys. Rev. B} \textbf{7}, 1920-1931 (1973)
\bibitem{Willnecker90}Willnecker, R. \textit{et al.} Dislocation velocities, dislocation densities, and plastic flow in lithium fluoride crystals. \textit{Appl. Phys. Lett.} \textbf{56}, 324-340 (1990).
\bibitem{Messerschmidt10}Messerschmidt, U. Dislocation Dynamics During Plastic Deformation, Springer-verlag, Berlin (2010).
\bibitem{Johnston59}Johnston, W. G. \textit{et al.} Dislocation velocities, dislocation densities, and plastic
flow in lithium fluoride crystals, \textit{J. Appl. Phys.} \textbf{30}, 129-145 (1959).

\end{thebibliography}

\end{document}